# Topologically Variable and Volumetric Morphing of 3D Architected Materials with Shape Locking

Kai Xiao[1], Yuhao Wang[1], Chao Song[2], Bihui Zou[1], Zihe Liang[1], Heeseung Han[1], Yilin Du[3], Hanqing Jiang[2],

Jaehyung Ju[1]*

**Abstract**: The morphing of 3D structures is suitable for i) future tunable material design for customizing material properties and ii) advanced manufacturing tools for fabricating 3D structures on a 2D plane. However, there is no inverse design method for topologically variable and volumetric morphing or morphing with shape locking, which limits practical engineering applications. In this study, we construct a general inverse design method for 3D architected materials for topologically variable and volumetric morphing, whose shapes are lockable in the morphed states, which can contribute to future tunable materials, design, and advanced manufacturing. Volumetric mapping of bistable unit cells onto any 3D morphing target geometry with kinematic and kinetic modifications can produce flat-foldable and volumetric morphing structures with shape-locking. This study presents a generalized inverse design method for 3D metamaterial morphing that can be used for structural applications with shape locking. Topologically variable morphing enables the manufacture of volumetric structures on a 2D plane, saving tremendous energy and materials compared with conventional 3D printing. Volumetric morphing can significantly expand the design space with tunable physical properties without limiting the selection of base materials.

## Introduction

Compared to conventional materials by phase change[1-3], reconfigurable metamaterials triggered by mechanical [4-9], thermal[10-14], magnetic[15,16], and electric[17,18] fields have significant potential for achieving tunable physical properties in intelligent material designs. However, there are few studies on reconfigurable metamaterials on the tunable range of physical properties owing to their inability to customize shape changes. Moreover, most reconfigurable structures have 1D or 2D shape changes[9,19-32], which limits the 3D tunable materials design. Therefore, three-dimensional (3D) morphing methods with dramatic volumetric changes and power-free locking methods for holding geometries are required for future tunable material designs.

Morphing from a 2D plane into a 3D volumetric shape can also significantly contribute to energy-efficient manufacturing by directly embedding intelligent designs into the fabrication process. Extra labor may not be required to assemble 3D shapes because of the single-step fabrication with morphing[23,28,33].

Furthermore, this topologically variable morphing does not require supporting materials for 3D printing to construct heterogeneous porous geometries[9,13,22,31], saving materials and reducing the energy crucial for fabrication by faster manufacturing owing to reduced fabrication volume without supporting materials[23,33]. Before morphing, the 2D plane shape contributes to efficient storage and transport while saving significant storage volume[13,23]. The morphing strategy can be extremely useful in micro- and nano-fabrication, where the direct fabrication of 3D complex geometries is challenging; 2D plane-based fabrication is more conventional, for example, electron beam nanolithography[34], dip-pen nanolithography[35], and direct-write atomic layer deposition[36].

For advanced tunable material designs, morphing between two or more target shapes requires an inverse design method. However, most inverse morphing designs are still in the 1D[25-27] and 2D stages[9,20-24,28-32,37-39]. Furthermore, low-dimensional morphing has limited structural applications in building programmable materials; morphed shells and plates do not provide sufficient stiffness in the lateral direction[11,19]. Therefore, the morphing of volumetric ligaments must be introduced advanced tunable material designs. Moreover, morphed shapes for structural applications must also be stable without additional holding energy. However, few studies on power-free shape locking of morphing structures[19] have been explored.

This study aimed to develop the topologically variable morphing of metamaterials from 2D to 3D geometries using an inverse design method with shape locking in orthogonal and curvilinear 3D space, and explores volumetric morphing to design extreme bulk and shear modulus tunability. Identifying the morphing conditions using the chosen unit cell types with instability can provide the main idea for designing and manufacturing topologically variable morphing devices. Our method can be used for tunable metamaterial design for a wide range of bulk and shear moduli in the automotive, aerospace, civil, and biomedical engineering fields.

## Topologically variable morphing

Most existing shape-morphing approaches transform 1D sticks into spatial 1D curves or 2D planes into spatial 2D surfaces[9,20-25,27-32,37-39]. These morphing strategies work in the same topology, characterized by the Euler equation $\chi = V - E + F = 1$, based on the number of vertices ($V$), edges ($E$), and faces ($F$) in the discretized target shapes, as illustrated in Fig. 1a, with varying mean and Gaussian curvatures. However, no studies have been conducted on topologically variable morphing (Fig. 1b). In addition, a volumetric morphing of $\chi = 0$ in Fig. 1e or $\chi = 2$ has not been explored yet. Limited knowledge of topologically variable and volumetric morphing limits the design space for morphing to transform highly complex 3D shapes and restricts programmable material tunability, such as practical 3D curvilinear geometries and closed surfaces with singularities. Therefore, we investigate reversible multi-dimensional morphing between plane and 3D shapes, including curvilinear geometries such as spheres, hyperboloids, cubes, cones, and combinations of those with singularities and sharp edges on vertices.

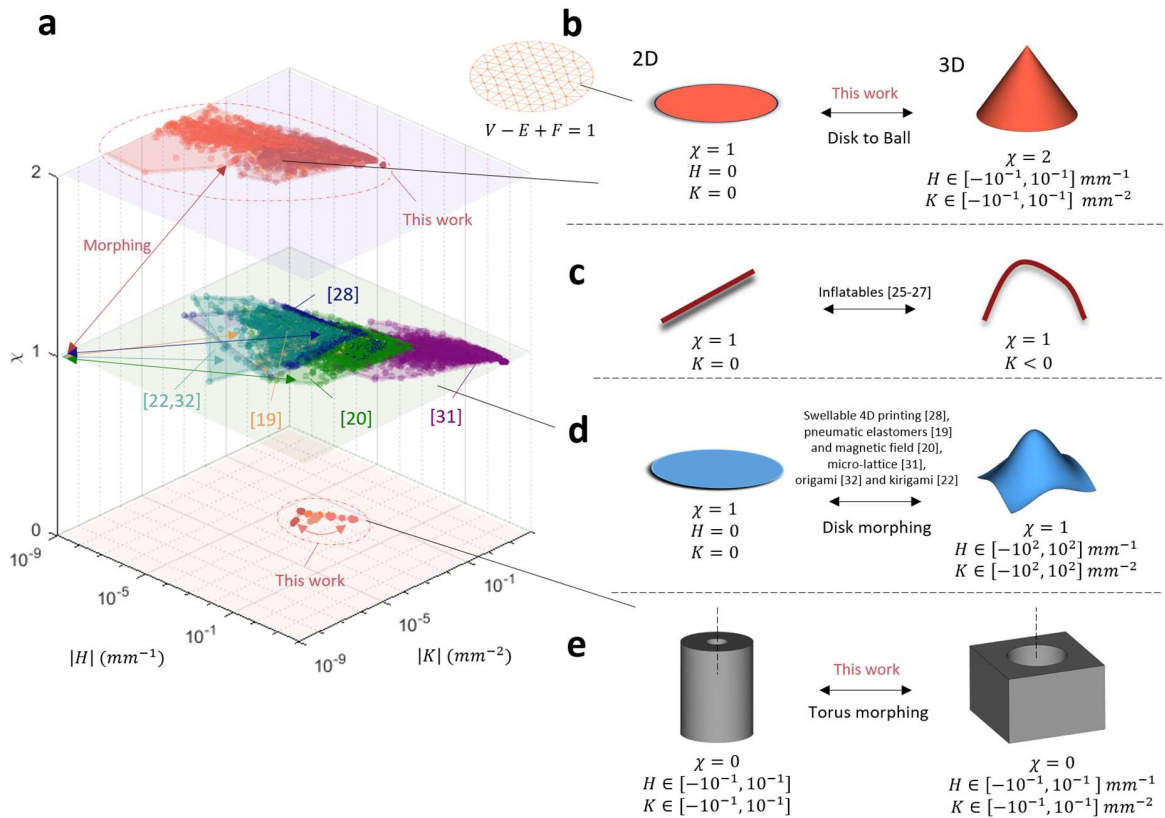

Fig. 1. (a) Design space for the morphing of structured materials; the design space is characterized by a topological parameter - Euler characteristic $\chi$ and two geometric parameters – mean $|H|$ and Gaussian $|K|$ curvatures. $\chi$ indicates the global topological property of shapes. $|H|$ and $|K|$ indicate the local geometric features. Higher values of $|H|$ and $|K|$ correspond to finer or smaller geometric features. A single shape corresponds to multiple scatters, and all these scatters are calculated based on the shapes at the achieved manufacturing scale. (b) Schematics of the topologically variable morphing from a disk to a cone. Euler characteristic $\chi = V - E + F$, where $V$ represents the number of vertices, $E$ is the number of edges, and $F$ is the number of faces for a meshed shape. Schematics of topologically invariable morphing between two (c) sticks, (d) disks, and (e) torus.

Fig. 2 illustrates the topologically variable morphing of architected materials from 2D flat surfaces to 3D curvilinear shapes. We selected a flat foldable unit, which is a Sarrus-linkage origami[6], discretized a 3D target shape, which is a sphere with volumetric mapping of the units in Fig. 2a. The square surface of the unit is triangulated and separated using a flexible hinge (Fig. 2b). The triangular faces functioned as stiff panels, and the edges between the faces served as hinges. Furthermore, we stacked the spatially periodic units into an assembly for a particular system size, such as $2 \times 2 \times 2$ in Fig. 2c, and mapped the assembly to a 3D curvilinear shape using optimal mass transport[40]. This volumetric mapping generated an initial discretization, where each unit cell heterogeneously deformed to fill the target volumetric shape, as shown in Fig. 2c. However, this initial discretization is not yet foldable. To enable foldability to a flat state, we implemented an inverse design using kinematic constraints, where the folding condition between the two states can be described by the following constraint:

$$L_{i_j}(\mathbf{v}_1, \mathbf{v}_2, \ldots, \mathbf{v}_V) = \bar{L}_{i_j}(\bar{\mathbf{v}}_1, \bar{\mathbf{v}}_2, \ldots, \bar{\mathbf{v}}_V) \qquad (1)$$

where $L_{i_j}$ in Figs. 2c.2 is the length of the $i$-th edge on the $j$-th face of the initially discretized configuration in 3D space, and $\bar{L}_{i_j}$ in Fig. 2d.2 is the length of the counterpart edge in the initial flat configuration in Fig. 2c.1. $\mathbf{v}$ and $\bar{\mathbf{v}}$ are the position vectors of the vertices of the expanded and flat configurations in the intermediate state, as shown in Figs. 2c.1 and 2d.1, respectively. $V$ is the total number of vertices in the structure.

Equation (1) ensures the existence of two morphable states – flat and expanded–in one architected material. Despite Equation (1) modifying the length of the initial discretization, we employ a boundary constraint on the initial expanded configuration to maintain the 3D target shape.

$$|\mathbf{v}_k - \widetilde{\mathbf{v}_k}| = 0 \qquad (2)$$

where $\mathbf{v}_k$ is the position vector of the $k$-th node in the initial expanded state. $\mathbf{v}_1, \mathbf{v}_2$ and $\mathbf{v}_3$ in Fig. 2c.1 must tie to the spherical boundary. $\widetilde{\mathbf{v}_k}$ associates the nearest vertices to the $k$-th node on the sampled surface of a 3D target shape, as shown in Fig. 2a.

Furthermore, we enforced a non-collision condition on the triangular segments of the expanded configuration by utilizing the separating axis theorem[41] to constrain the overlapping scalar $O_s$ to a negative value, ensuring that the chosen pairs of triangular segments do not intersect. This constraint can be expressed as follows:

$$O_s(\mathbf{v}_1, \mathbf{v}_2, \ldots, \mathbf{v}_V) < 0 \qquad (3)$$

where $O_s$ is the overlapping scalar of the $s$-th combination of the two triangular faces. More details about selecting the pairs of triangular segments and calculating $O_s$ in Section 1.1 in the Supplementary Information. An objective function searches for the optimum morphing structures, while minimizing the positional alteration of the vertices relative to the initial discretization. Therefore, we solve this function as follows:

$$\min \sum_{v=1}^{V} \left|\mathbf{v}_{v_f} - \mathbf{v}_v\right|^2 \qquad (4)$$

$\mathbf{v}_{v_f}$ represents the $v$-th vertex of a foldable structure, as shown in Fig. 2e, and $\mathbf{v}_v$ is associated with the corresponding vertex in the initial discretization, such as $\mathbf{v}_1, \mathbf{v}_2$ and $\mathbf{v}_3$ in Fig. 2c.

Using the inverse design framework of Equations (1)–(4), Figs. 2e-2h demonstrate the topologically variable morphing for different targets of 3D curvilinear shapes and system sizes. Moreover, our design strategy can be applied to other units (e.g., stacked Mirua units), as presented in Supplementary Section 1.2 and Fig. S2, demonstrating the generality of topologically variable morphing. Supplementary Videos 1 and 2 demonstrate reversible and topologically variable morphing between $\chi = 1$ and $\chi = 2$ in Figs. 2f-h.

Further details regarding the boundary conditions, manufacturing, and experimental comparison are provided in Supplementary Section 2.1.

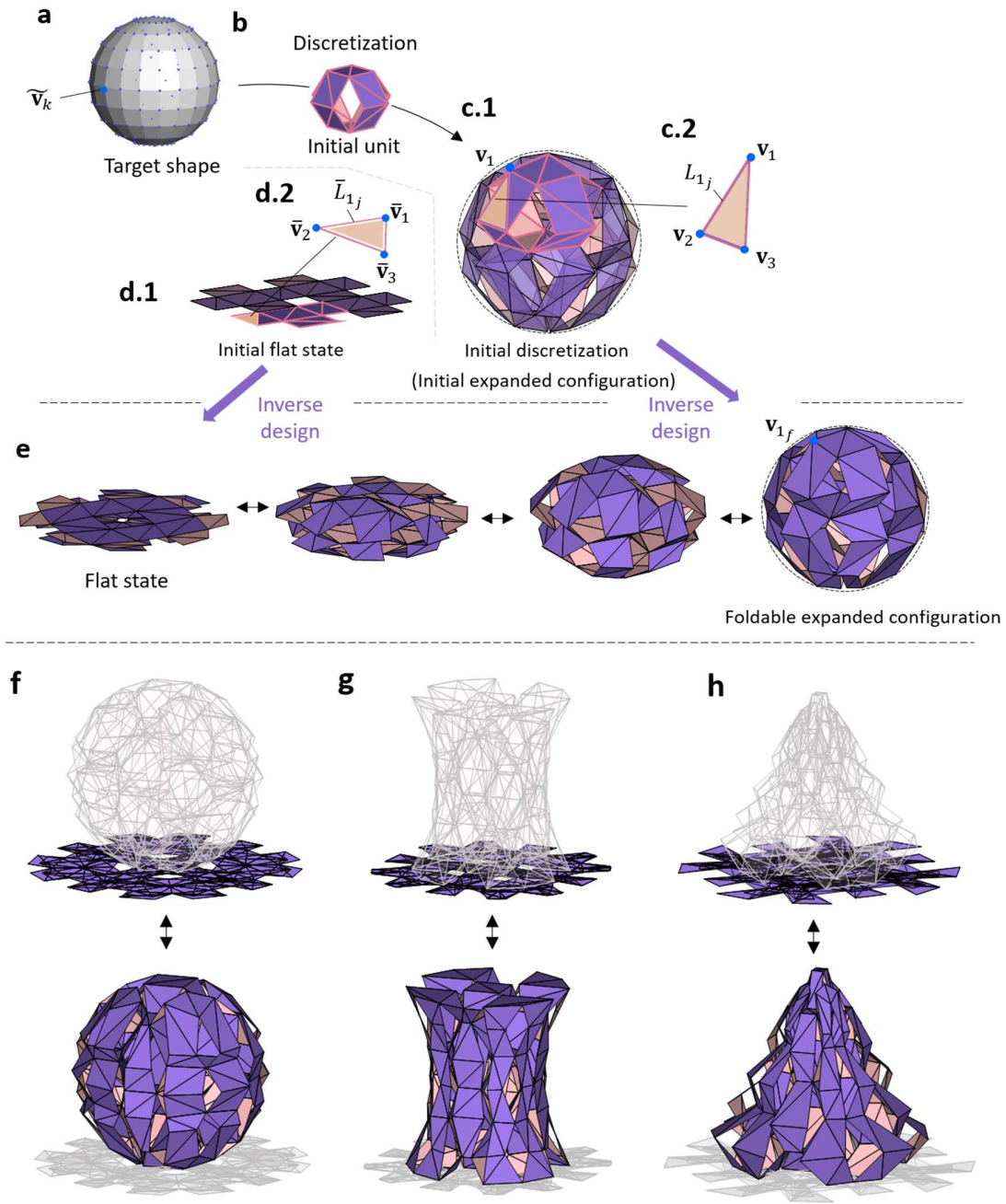

*Fig. 2. Topologically variable morphing of architected materials from 2D flat surface to 3D curvilinear shapes. (a) A spherical morphing target. (b) Discretization of a sphere with a Sarrus-linkage unit[6] for a given system size, e.g., $2 \times 2 \times 2$. The square surface of the unit is triangulated and separated by a flexible hinge. (c.1) the initial discretization of a sphere with $2 \times 2 \times 2$ units. (c.2) A triangular segment of the initial discretization in the sphere. (d.1) The preliminary flat state obtained by regular units' projection on a 2D plane. (d.2) A triangular segment in the initial flat state. (e) Topologically variable morphing of modular origami structures from a 2D flat state to a sphere one. (f-h) Other examples of topologically variable morphing with a $3 \times 3 \times 3$ system size for target 3D curvilinear shapes.*

## Morphing with shape locking

The addition of a locking functionality to topologically variable morphing adds more value to energy-efficient manufacturing, which is significant for structural applications after deployment. Rather than using a material's freezing at the molecular level, such as shape memory polymer shape fixity[10,11,42], which limits the selection of materials, we used structural instability[43]. By exploring the potential lockable features of origami structures, we implemented bistability to design the topologically variable morphing of 3D curvilinear modular origami structures, producing stable energy states in the initial and deployed configurations. The integration of the morphing algorithm with bistable unit cells enables shape-lockable morphing. To verify the bistable morphing of the architected materials, we implement a numerical simulation with a non-rigid panel assumption, and construct the target structures with an assembly of transformed unit cells divided by triangle elements consisting of bars and hinges. We obtained this nonlinear mechanical deformation during deployment using the software – Merlin2[44,45], which provides force-displacement behavior and energy variation as a deployment function.

Topologically variable morphing with shape locking is significant for practical applications; moreover, a biomedical stent must maintain its expanded shape to withstand pressure from the vessel walls, ensuring consistent blood flow[46]. A space telescope must precisely unfold and align its mirror segments to form a single large mirror assembly, which requires stability[39,47]. The flat and expanded states can maintain shapes with low potential energies, while the other intermediate states maintain relatively high energies during the morphing process.

Based on previous studies[48], the geometry of the modular origami unit in Fig. 3a can be modified to achieve alternative flat-foldable morphing with shape locking, enabling the inward folding of its faces (Fig. 3b). A tilting angle $\theta$ in Figs. 3a-3c characterizes the bistability. Furthermore, we define a parameter $\omega = (\mathbf{p}_2 \times \mathbf{p}_1) \cdot \mathbf{p}_3$, which differentiates between outward ($\omega > 0$) and inward ($\omega < 0$) configurations. In addition to $\omega$ and $\theta$, the difference in faces' height, $\delta$, is also a crucial parameter to influence actuation energy. Additionally, we define $\delta = d_2 - d_1$ and $\delta = d_4 - d_3$, while maintaining $d_1 = d_3$ and $d_2 = d_4$ in the periodic unit cell $a, b,$ and $c$. The parameter $d$ denotes the peak facet height during morphing, as shown in Figs. 3a-3c.

The force-displacement and energy wall diagrams in Figs. 3d and 3e validate the selection of $\omega, \theta$ and $\delta$ as the design parameter for tuning the actuation energy. Compared with the original unit ($\theta = 0°, \omega > 0$ and $\delta = 0$), the unit whose $\theta = 30°$, $\omega < 0$ and $\delta = 0.1\, d_{max}$, where $d_{max} = \max(d_1, d_2, d_3, d_4)$, require a significant force and actuation energy for morphing, while maintaining a low energy state (i.e., stability) in both the 2D flat ($\chi = 1$) and 3D volumetrically expanded ($\chi = 2$) configurations. The following constraints provide bistable periodic units.

$$\omega < 0 \tag{5}$$

$$\delta = d_2 - d_1 \neq 0 \tag{6}$$

$$\delta = d_2 - d_1 \neq 0 \tag{7}$$

$$d_1 = d_3 \tag{8}$$

$$d_2 = d_4 \tag{9}$$

For the construction of a volumetric shape with the volumetric mapping of unit cells, Equations (5)-(9) can be modified for the topologically variable morphing between 2D flat ($\chi = 1$) and 3D curvilinear ($\chi = 2$) assemblies:

$$\omega_{m_n} < 0 \tag{10}$$

$$d_{2_n} - d_{1_n} - \Delta d_{21_n} = \delta \tag{11}$$

$$d_{4_n} - d_{3_n} - \Delta d_{43_n} = \delta \tag{12}$$

$$d_{3_n} - d_{1_n} = \Delta d_{31_n} \tag{13}$$

$$d_{4_n} - d_{2_n} = \Delta d_{42_n} \tag{14}$$

where $m = 1,2,3,4$, associating with the $m$-th face folding inward or outward in the $n$-th unit in the assembly. $\Delta d_{m_1 m_{2_n}}$ is the initial face height difference for $m_1$-th and $m_2$-th faces of the nonperiodic units in the initial discretization after volumetric mapping.

By incorporating Equations (10)–(14) into the inverse design algorithm in Equations (1)–(4), we obtain the actuation energy, which is denoted as $E_{max}$ in a $2 \times 2 \times 2$ spherical structure. $E_{max}$ is the maximum stored energy during morphing and relies on the values of $\delta$ and $\theta$, as shown in Fig. 3g. Fig. 3g shows a design map of the morphing energy $E_{max}$ of a spherical assembly for varying geometric parameters. $D_{max}$ represents the maximum length of the structure. In Figs. 3f-3h, the diameter of the sphere $D_{max} = 200mm$.

Additionally, we plot the region of $\omega < 0$ and $\omega > 0$. Our findings indicate that lockable configurations can be more readily identified within the $\omega < 0$ region, where $E_{Max}$ and $|\delta|$ tend to exhibit a higher value. For example, we identify the configurations on the border of the $\omega < 0$ region in Fig. 3g, which necessitates over one-hundred times the energy required for actuation compared to the region of $\omega > 0$ (Fig. 3h). Furthermore, we distinguished the bistable region from the monostable region using the criterion of $\frac{E_{Max}}{E_{Hinge}} = 10$. The bistable region in Fig. 3g demonstrates snap-through behavior, characterized by a sudden jump in the negative force during the folding process; further details and a comparison of experimental observations can be found in Supplementary Section 2.1, and Supplementary Video 3. Our topologically variable morphing with shape locking can be applied to various geometric parameters, such as the selection of unit cell geometry, number of unit cells, and morphing target shapes. Additional results are presented in Supplementary Section 2.2.

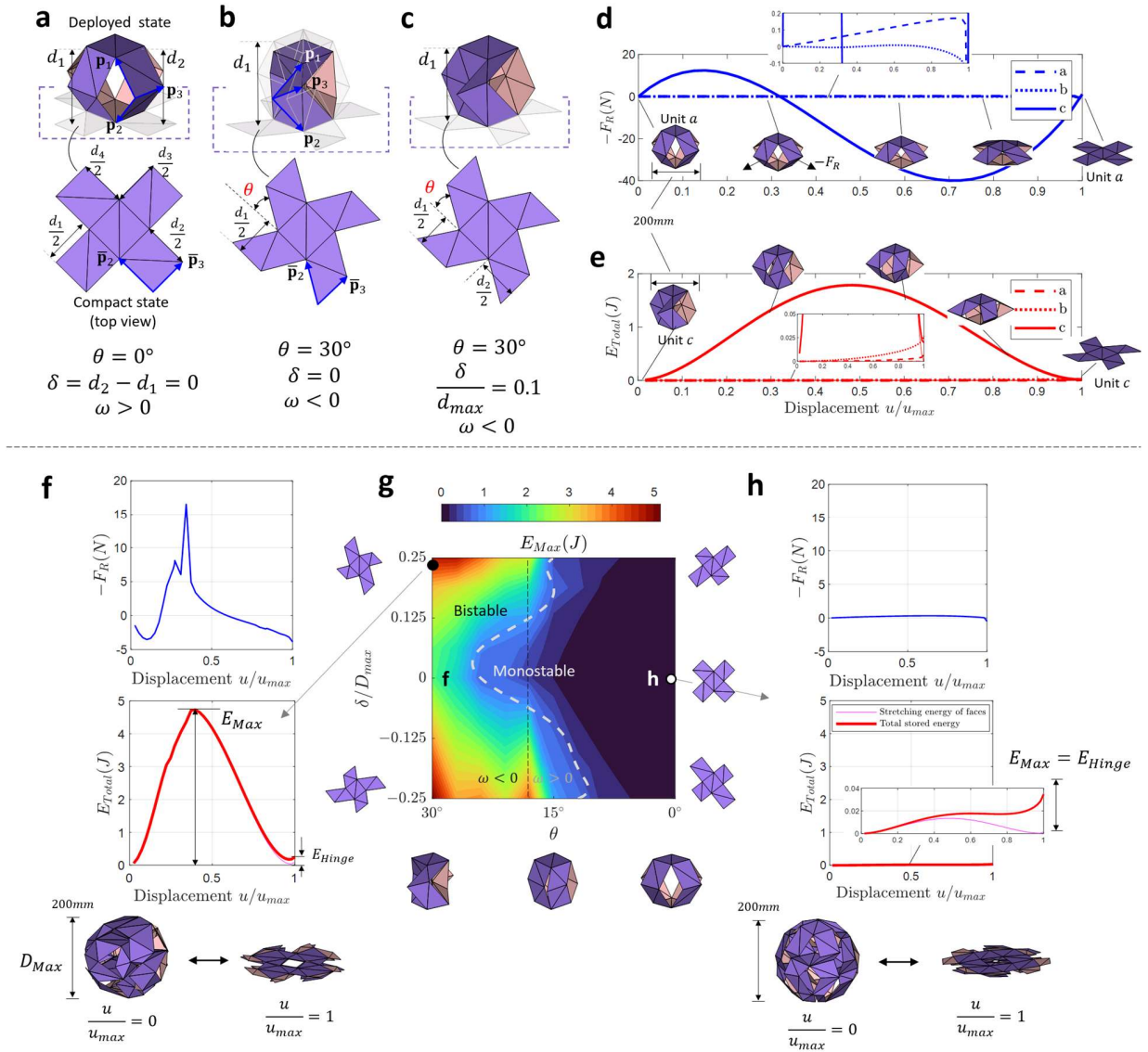

Fig. 3. ***Topologically variable and shape lockable morphing of 3D curvilinear modular origami structures:*** *(a) The unit cell with $\theta = 0°, \delta = 0, \omega > 0$ (b) The unit cell whose faces are snapped inwardly, where $\theta = 30°, \delta = 0, \omega < 0$. (c) The unit with inwardly snapped faces and non-uniform expansion height, where $\theta = 30°, \delta = 0.1\, d_{max}, \omega < 0$, where $d_{max} = max\,(d_1, d_2, d_3, d_4)$. (d) The force-displacement curve of the units $a, b$ and $c$ during the topologically variable morphing (e) Stored energy $E_{total}$ in the unit $a, b$ and $c$ during the morphing process. (f) The mechanical response of a $2 \times 2 \times 2$ spherical assembly of unit 'b' demonstrates high actuation energy with bistability. (g) A design map of $2 \times 2 \times 2$ spherical assembly with actuation energy $E_{Max}$. The bistable and monostable regions are divided by quantifying $\frac{E_{Max}}{E_{Hinge}} = 10$. (h) A spherical assembly of unit 'a' shows low actuation energy and no bistability.*

## Volumetric morphing for materials design

In addition to the topologically variable morphing ($\chi = 1 \leftrightarrow \chi = 2$), a volumetric morphing with $\chi = 0$ is also significant in the 3D space material design. To implement volumetric morphing, we use the curvilinear shape of unit cell $b$ in Fig. 3b to discretize two target shapes – a hollow cylinder and a hollow cube with $\chi = 0$, as shown in Fig. 4a. Furthermore, the curvilinear unit $b$ in Fig. 4b stacks along the circumferential, radial, and vertical directions, limiting flat foldability owing to the symmetry breaking of the geometry of the curvilinear units. However, the curvilinear unit $b$ can contribute to the generalized volumetric morphing of complex 3D curvilinear shapes.

The volumetric tessellation of the curvilinear unit $b$ (State #1) in Fig. 4b can construct a discretized cylinder with a system size of $12 \times 2 \times 3$, as shown in Fig. 4c.1. In parallel, cylindrical stacking and volumetric mapping of another curvilinear unit $b$ (state #2) to a hollow cuboid can provide a discretized cuboid with a system size of $12 \times 2 \times 3$, as shown in Fig. 4c.2. More details regarding the discretization are provided in Supplementary Section 1.3. The inverse design algorithm conducts volumetric morphing by adjusting the initial discretized geometries (hollow cylinder and hollow cuboid) with foldability, non-overlapping, and boundary constraints. Volumetric morphing provided tremendous tunability in the second area moment of inertia between the two target morphing shapes: $I_d^1/I_c^1 = 18.6$ with cross-sectional area changes of $A_d^1/A_c^1 = 3.6$ in Figs. 4d and 4e. Moreover, one can morph between two hollow cylinders, as shown in Fig. 4f, which provides a different tunable range for the second area moment of inertia for a structural beam design: $I_d^2/I_c^2 = 10.5$ with $A_d^2/A_c^2 = 2.7$. The effective cross-sectional area and second area moment of inertia of the geometries in Figs. 4e.1 and 4f.1 are the same: $A_c^1 = A_c^2$ and $I_c^1 = I_c^2$. However, morphing into different shapes, for example, one for a cuboid and the other for a cylinder, can provide different structural beam properties - $I_d^1 = 1.7\ I_d^2$ and $A_d^1 = 1.3\ A_d^2$, as shown in Figs. 4e.2, 4f.2, and 4j.1. Therefore, the bending stiffness of metamaterials with morphing geometries can be customized.

Furthermore, we implemented the sphere-cube (Fig. 4g) and sphere-sphere morphing (Fig. 4h) with a similar spherical discretization technique using the same system size of $10 \times 4 \times 1$ along the azimuthal, polar, and radial directions, respectively.(See more details regarding the spherical discretization in Supplementary Section 1.3). Using homogenization with FE software (Merlin2)[44] we identified the tunable range of the effective bulk and shear moduli by applying loading conditions for volumetric and shear deformations, as shown in Fig. 4i. The sphere-cube morphing in Fig. 4g produces high incompressibility with low rigidity, indicating a high resistance to volume change and easy shape changes, as shown in Fig. 4j.2. Conversely, the sphere-sphere morphing in Fig. 4h provides a relatively low bulk modulus and high shear modulus compared with the sphere-cube morphing, as shown in Fig. 4j.2. The bulk-to-shear modulus ratio, $B/G$ by selecting the morphing mode, as shown in Fig. 4j.3. The global volumes in Figs. 4g.1 and 4h.1 are the same, but the internal structures are different, producing different mechanical properties. Furthermore, sphere-cube morphing produces a high bulk-to-shear modulus ($B/G \sim 600$), comparable to that of pentamode structures. Pentamode structures have fluid-like properties and cannot be used for structural applications.

Compared with existing natural and structured materials[49-52], our morphing metamaterials provide a new territory of material properties: tunable shear and bulk moduli - 0.01 to 0.3 MPa for shear modulus and 0.3 to 10 MPa for bulk modulus when the metamaterials design is conducted with polymers, as shown in Fig. 4k. Such tunable properties can be deliberately shifted or expanded by selecting other base materials, e.g., aluminum and steel. Elastomers and rubbers have low shear moduli and high bulk modulus and are incompressibility[53]. Our metamaterials made of polymers and metals can produce a lower shear modulus, indicating an easy shape change by an external load, as shown in Fig. 4k.1. Polymer foams exhibit low bulk moduli. However, our metamaterials provide a lower bulk modulus, as shown in Fig. 4k.2.

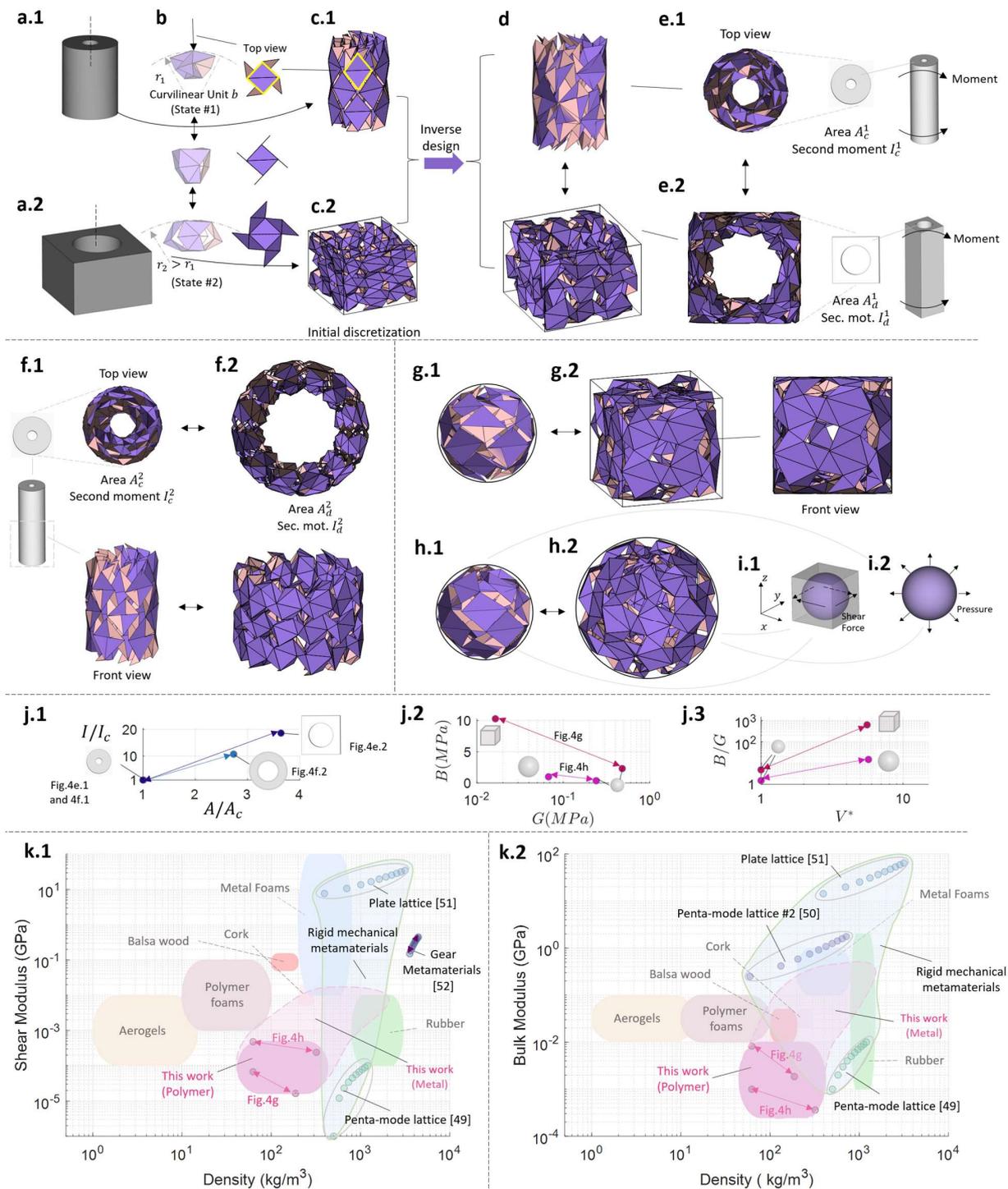

*Fig. 4. Volumetric morphing of architected materials with shape locking. (a) Two morphing target shapes with $\chi = 0$; (b) The corresponding curvilinear shapes of unit b for volumetric morphing; (c) Volumetric mapping of curvilinear shapes of unit b (States #1 and #2) into a hollow cylinder and a hollow cuboid, respectively; (d) The optimized morphing structures that can reversibly morph with shape locking from hollow cylindrical to hollow cuboidal shapes; (e) The top view of the optimized morphing structures; (f) A cylinder-cylinder morphing; (g) A sphere-sphere morphing; (h) A sphere-cube morphing; (i) Schematics of shear and bulk resistance of a morphable sphere; (j.1) The second area moment inertia ratio $I/I_c$ for the ratio of cross-sectional areas $A/A_c$; (j.2) The bulk (B) and shear (G) moduli of morphing structures based on a polymer (polylactic acid). (j.3) The bulk to shear moduli ratio*

$(B/G)$ for normalized volume $V^*(=V/V_c)$; newly occupied material design space in shear (k.1) and bulk (k.2) Moduli of the morphed structures, compared with existing natural materials and metamaterials[49-52]. See the morphing process of the architected materials in Supplementary Video 4.

## Topologically variable morphing of complex shapes

To demonstrate the applicability of our design strategy in practical engineering or artistic structures, architecture, and furniture, whose shapes are concave and involve singularities with sharp edges and points, we showcase examples of the topologically variable morphing of airplanes, lion sculptures, and sofa chairs. The discretization process of these shapes involves the alignment of the assembled initial units to concave forms and the volumetric mapping of the assembly to the target morphing geometry. After implementing the inverse design algorithm of Equations (1)–(4), Figures 5a-5c show the flat morphing of the complex shapes.

The complexities of the morphed structures were quantified by plotting the mean $|H|$ and Gaussian $|K|$ curvatures of the morphed geometries on the plane of $\chi = 2$, as shown in Fig. 5d. Compared with simple shapes, such as spheres, cones, and hyperboloids in Figs. 2f-2h, whose curvatures are concentrated in a single point set on the curvature map in Fig. 5e, complex shapes with severely varying curvatures occupy a much larger area, as shown in Fig. 5d. These curvatures demonstrate the broad adaptability of our inverse design method, which applies to the morphing of ubiquitous structures.

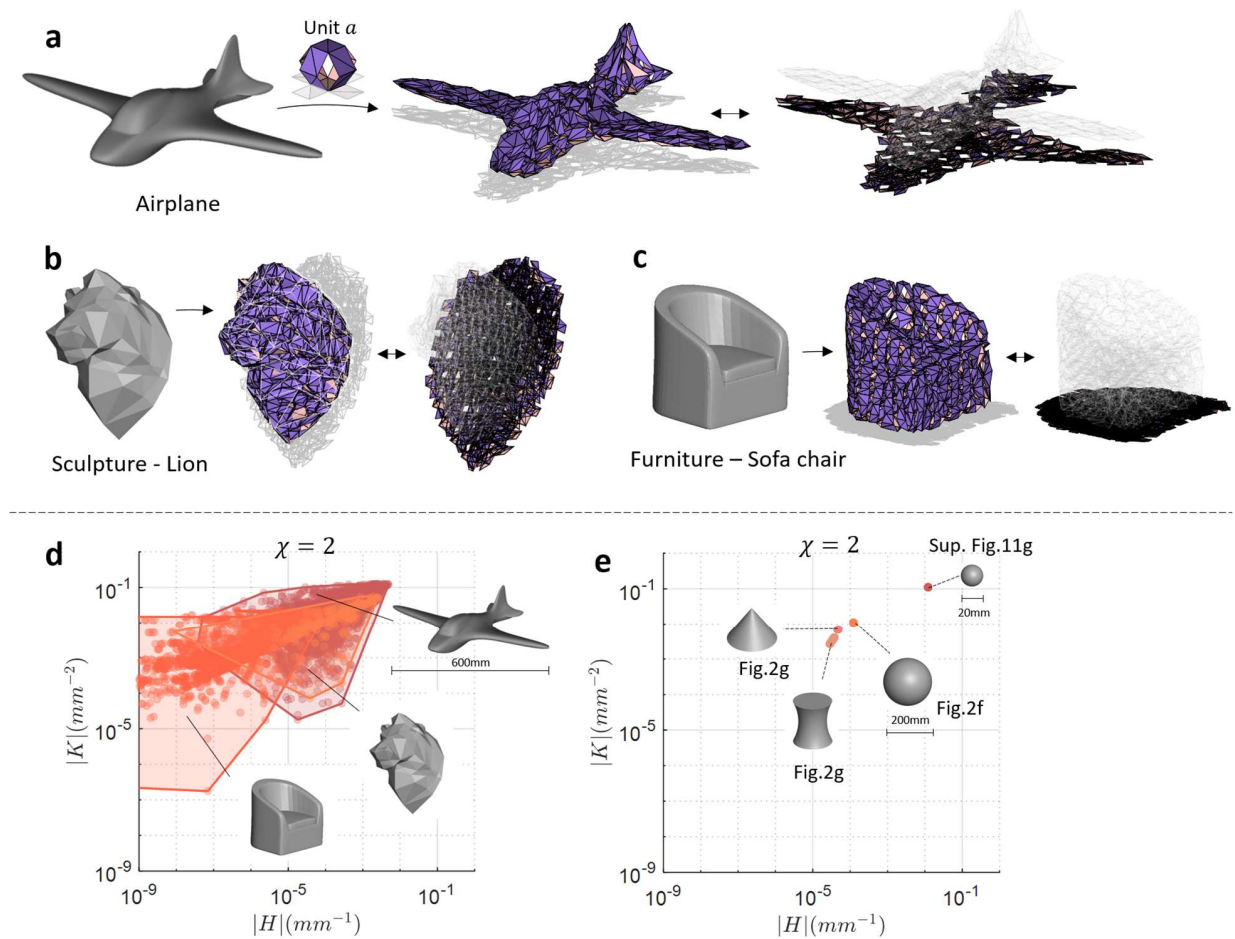

Fig. 5. Topologically variable morphing of complex geometries; (a) an engineering structure – an airplane, (b) an artistic sculpture of a lion, and (c) furniture - a sofa chair; (d) Mean $|H|$ and Gaussian $|K|$ curvatures of the meshed shapes (airplane, lion and sofa

*chair), Note each shape corresponds to multiple scatters with varying |H| and |K| bounded by a colored solid line. (e) Mean |H| and Gaussian |K| curvatures of the simple shapes (sphere, hyperboloid, and cone).*

## Discussion

Morphing is a significant tool for designing tunable materials and structures to customize their properties. However, no inverse design method for morphing is available for practical engineering applications, such as topologically variable and volumetric morphing with shape locking. This study demonstrates that the volumetric mapping of any bistable unit cell with kinematic and kinetic adjustments between morphing targets can produce flat-foldable and volumetric morphing structures. Furthermore, the morphed shapes retain high stiffness simultaneously owing to their mechanical instability.

Previous morphing strategies were based on an inverse design for morphing between spatial curves or surfaces[9,20-25,27-32,37-39], limiting their practical applications. The inverse design in this study advances the morphing of ubiquitous 3D structures, including complex curvilinear shapes[54]. Shape locking is a powerful tool for retaining stiffness in metamaterial design of metamaterials[6,48,55]. However, it can be empowered by the inverse design of morphing to enlarge the tunable ranges of the metamaterial properties and to design the bulk and shear moduli, as demonstrated in this study. Previously, there have been two types of design efforts for the bulk and shear moduli. First, the Poisson's ratio, the negative Poisson's ratio of an isotropic material system provides a high shear modulus[56], where shear and bulk moduli designs highly depend on the Poisson's ratio. Pentamode structures, called meta-fluids, have high bulk-to-shear modulus ratios ($10,000 > B/G > 100$) and cannot be used for structural applications[49,57]. Natural gold has a high bulk-to-shear modulus ratio ($B/G \sim 13$)[53]. The morphed structures in Fig. 4g exhibit a significantly high bulk-to-shear modulus ($B/G \sim 600$) with structural potential by shape locking. Morphable second-area moments of inertia can create turnability in structural engineering designs. The same initially designed cross-sectional area can produce a different second-area moment of inertia, as shown in Fig. 4j.1. Furthermore, bending is the dominant loading mode in structural applications[53], which our morphing strategy has a broad influence.

The topologically variable morphing of flat foldable structures can provide tremendous potential for saving energy and materials during fabrication compared to conventional 3D and 4D printing[28]. The decoupling geometry for fabrication and deployment in topologically variable morphing can significantly impact the logistic control of engineering structures through the easy storage and transportation of flat-foldable morphing structures. This topologically variable morphing can highly influence micro- and nano-fabrication, where the direct fabrication of 3D complex structures remains challenging.

This study can be used directly for actuators that can be applied to intelligent morphing metamaterials and soft robots because the potential actuation is demonstrated with magnetic control in Supplementary Video 5. Integrating our morphing strategy with mechanoelectric-coupled materials, such as piezoelectric materials[18,58] and dielectric polymers[59] can advance robotic metamaterial sensing by producing electrical signals with customized mechanical deformation. Morphing with shape-locking can enlarge the design space of mechanical logic gates for mechanical computing. Moreover, there is limited geometry in logic gate design for mechanical computing, such as the Kresling[60] and Miura-ori geometries[61].

## Conclusion

This study demonstrates that the volumetric mapping of non-periodic unit cells with kinematic (foldability) and kinetic (bistability) modifications between morphing targets can produce a generalized inverse design method for the topologically variable and volumetric morphing of metamaterials with shape locking. Volumetric morphing combined with bistability creates new tunable territories of material and structural properties, such as bulk/shear moduli and the second area moment of inertia. Topologically variable morphing integrated with bistability can synergistically advance the manufacturing of 3D complex structures with vast savings in materials, time, and energy for fabrication and volume-saving logistic control. The inverse design method for 3D complex shapes with shape locking can contribute to the design of next-generation engineering morphing structures for automotive, aerospace, civil, and biomedical engineering applications. Integrating the morphing strategy with external environments such as light, thermal, electric, and magnetic fields can advance robotic and physical intelligence, including advanced motion control of robotic metamaterials for sensing and actuation and 3D morphable logic gates with instability for mechanical computing.

## Acknowledgements

This research is supported by the National Natural Science Foundation of China (Grants no. 12272225), the Ministry of Science and Technology in China (Grant no. SQ2022YFE010363), and the Research Incentive Program of Recruited Non-Chinese Foreign Faculty by Shanghai Jiao Tong University.